\documentclass[modern]{aastex62}

\shorttitle{Different contributions to space weather and space climate}
\shortauthors{Jie Jiang et al.}

\begin{document}

\title{Different contributions to space weather and space climate from different big solar active regions}

\correspondingauthor{Jie Jiang}
\email{jiejiang@buaa.edu.cn}

\author[0000-0001-5002-0577]{Jie Jiang}
\affil{School of Space and Environment, Beihang University, Beijing, China}

\author{Qiao Song}
\affiliation{Key Laboratory of Space Weather, National Center for Space Weather, China Meteorological Administration, Beijing 100081, China}

\author{Jing-Xiu Wang}
\affiliation{Key Laboratory of Solar Activity, National Astronomical Observatories, Chinese Academy of Sciences, Beijing 100012, China}

\author{T\"{u}nde Baranyi}
\affiliation{Debrecen Heliophysical Observatory (DHO), Konkoly Observatory, Research Centre for Astronomy and Earth Sciences, Hungarian Academy of Sciences, 4010 Debrecen, P.O. Box 30, Hungary}



\begin{abstract}
The purpose of this paper is to show that large active regions (ARs) with different magnetic configurations have different contributions to  short-term and long-term variations of the Sun. As a case study, the complex $\delta$-type AR 12673 and the simple $\beta$-type AR 12674 are investigated in detail. Since the axial dipole moment at cycle minimum determines the amplitude of the subsequent cycle and space climate, we have assimilated the individual observed magnetic configurations of these two ARs into a surface flux transport model to compare their contributions to the axial dipole moment $D$. We find that AR 12673 has a significant effect on $D$ at the end of the cycle, making it weaker because of the abnormal and complicated magnetic polarities. An initial strongly positive $D$ ends up with a strongly negative value. The flare-poor AR 12674 has a greater contribution to the long-term axial dipole moment than the flare-rich AR 12673. We then carry out a statistical analysis of ARs larger than 800 $\mu$Hem from 1976 to 2017. We use the flare index FI and define an axial dipole moment index DI to quantify the effects of each AR on space weather and space climate, respectively. Whereas the FI has a strong dependence on the magnetic configuration, the DI shows no such dependence. The DI is mainly determined by the latitudinal location and the latitudinal separation of the positive and negative magnetic fluxes of the ARs. Simple ARs have the same possibility as complex ARs to produce big DI values affecting space climate.
\end{abstract}

\keywords{Sun: magnetic fields, Sun: activity}


\section{Introduction} \label{sec:intro}
The Sun's magnetic activity varies over different time scales. Short-term variations in different forms of solar activity, e.g., solar flare and coronal mass ejection, and their effects on the near-Earth environment and technology are usually referred to as space weather. Long-term variations, e.g., those longer than years, in solar activity and their effects are usually referred to as space climate \citep{Mursula2007,Nandy2007}.

Solar flares are one of the main sources of disastrous space weather events. They are thought to be caused by the release of magnetic energy \citep[see][and references therein]{Chen2011, Shibata2011}. It has been known for half a century that active regions (ARs) with more complex magnetic configurations and larger areas tend to produce flares of greater magnitudes \citep{Kunzel1960,Zirin1987,Sammis2000}. The Mount Wilson classification system for sunspot groups is widely used to describe the magnetic complexity of ARs. The system consists of four parameters in ascending order of the magnetic complexity, i.e., $\alpha$ (unipolar), $\beta$ (bipolar), $\gamma$ (multipolar), and $\delta$. The parameter $\delta$ is assigned to complex regions where at least one sunspot in the region contains opposite magnetic polarities inside of a common penumbra separated by no more than 2$^\circ$ in heliographic distance. Although the $\delta$-type ARs are approximately 5\% of all ARs \citep{Jaeggli2016}, different statistical samples show that more than 80\% of X-class flares are from ARs exhibiting $\delta$ sunspots \citep[e.g.,][]{Zirin1987, Shi1994, Guo2014, Toriumi2017}. The properties of $\delta$-type ARs that host flaring activity have been investigated extensively in previous studies \citep[e.g.,][]{Liu2002, Jing2006, YangY2017}.

However, as suggested by \cite{Chen2012}, in each solar cycle there are some ARs that have very large areas (e.g., $\geq$ 1000 $\mu$Hem, i.e., millionths of the solar hemisphere), but do not produce any flares higher than the M1.0 class. Only about 20\% of all regions having area greater than 1000 $\mu$Hem produce flares over X1 class \citep{Sammis2000}.  Flare-poor, big ARs usually belong to the simple $\beta$ type and are ignored by the space weather community. The big AR 12674, which appeared on the solar disk in 2017 September, is a typical example in this regard. It is a regular $\beta$-type AR. The largest flare produced by AR 12674 is C5.2. By contrast, the quasi-simultaneous emergence of AR 12673 is a regular $\beta\gamma\delta$-type AR. It produced 27 M-class and 4 X-class flares, one of which is the largest one (X9.3) in solar cycle 24. To date, there are several publications that have studied the flare-productive AR 12673 \citep[e.g.,][]{YangS2017, Yan2018, Seaton2018, Shen2018}. To our knowledge, there have been no such studies on the flare-quiet AR 12674. Actually, these two big ARs, 12673 and 12674, have the same potential to affect the solar cycle evolution, and hence to impose a similar influence on space climate.

Within the Babcock-Leighton (BL) framework for solar dynamos \citep{Babcock1961,Leighton1969}, the large-scale radial field over the solar surface, especially the polar field or the axial dipole moment, is source of the toroidal field
responsible for solar magnetic activity in the subsequent cycle. \citep{Jiang2013, Cameron2015}. After tilted bipolar or multipolar magnetic regions emerge on the solar surface, they have an initial axial dipole moment determined by their area and tilt angle. Flux transport processes over the surface have different effects on the axial dipole moment depending on the latitudes of the ARs' emergence. A lower latitude emergence generates a higher final axial dipole moment, due to the larger amount of cross-equator flux. \cite{Jiang2014b} demonstrated that the latitudinal dependence of the final axial dipole moment obeys a Gaussian function; see also \cite{Wang1991} and  \cite{Whitbread2018}. Since the total flux over the polar region is roughly equivalent to the flux of a large AR, the large ARs with high tilt angles and normal (abnormal) polarities emerging at low latitudes would have significant positive (negative) contributions to the axial dipole moment at cycle minimum and hence to the subsequent solar cycle. Here, normal (abnormal) polarities of ARs denote Hale ARs with their leading pole located equatorward (poleward) of their trailing pole and the anti-Hale ARs with their leading pole located poleward (equatorward) of their trailing pole. Evidences for the significant effect of a single special AR on the solar cycle are as follows.

\cite{Wang1991} and \cite{Whitbread2018} found that about half of the axial dipole moment at cycle minimum comes from about 10\% of ARs. \cite{Jiang2015} identified that the deep cycle 23 minimum and the weak cycle 24 are caused by a small number of large and abnormally oriented ARs. \cite{Nagy2017} demonstrated that large `rogue' regions can drastically affect the evolution of future solar cycles. Such large regions emerging during the early phases of a cycle can even affect the amplitude and duration of the same cycle. The effect of a single AR emerging in later phases can persist for multiple cycles. All of these studies indicated that large individual active regions with atypical properties can have a significant impact on the long-term behavior of the solar cycle. Studies have shown that different forms of solar activity, e.g., flare and CME occurrence rate and interplanetary shocks, track the solar cycle in both amplitude and phase \citep{Gopalswamy2004,Gopalswamy2006,Kilpua2015}. The long-term variation of solar activity corresponds to space climate. Throughout the paper, we simplify our concept as follows: the short-term variation of solar activity due to flares is regarded as the effects of ARs on space weather, and the long-term variation of solar activity produced by variation of the axial dipole moment in the time scale of solar cycle is regarded as the effects of ARs on space climate. Unlike the effects of ARs on space weather, which depend on the magnetic complexity and amount of flux of ARs, the effects of ARs on space climate depend on the latitudinal location and initial latitudinal separation of ARs. Therefore, flare-poor big ARs with simple magnetic configurations could also have significant effects on solar cycle variation and hence on space climate. This paper aims to provide a systematic demonstration of the different contributions from different types of big ARs to space weather and space climate.

The organization of this paper is as follows. In section \ref{sec:2ARs}, we investigate the different contributions to space weather and especially to space climate from the big ARs 12673 and 12674 in detail. A statistical study of big ARs from 1976 to 2017 is presented in Section \ref{sec:Statistics}. Finally, the summary and discussion are given in Section \ref{sec:summary}.

\section{Differences between two great Active Regions: AR 12673 and AR 12674} \label{sec:2ARs}
\subsection{General information on the two ARs}

The $\beta\gamma\delta$-type AR 12673 passed across the visible solar disk from 2017 August 28 to 2017 September 10. Based on the Debrecen Photoheliographic Data (DPD) sunspot catalogue \footnote{http://fenyi.solarobs.csfk.mta.hu/en/databases/DPD/}, this sunspot reached a maximum area $A$ of 1397 $\mu$Hem on September 7th with the latitudinal center $\lambda$ at 9$^\circ$.36 of the southern hemisphere. The tilt angle, which is measured based on the white-light image by the DPD, shows a large variation. The standard deviation of its tilt angle $\sigma_\alpha$ is 43$^\circ$.16. This standard deviation is calculated during the interval of August 31st to September 8th when its longitudinal distance from the central meridian (LDCM) is less than 60$^\circ$. The average tilt $\overline{\alpha}$ is 7$^\circ$.  AR 12674 is a standard $\beta$-type AR that occurred on the visible solar disk one day after AR 12673.  Based on the DPD catalogue, its area $A$ was 1238 $\mu$Hem on September 3rd with the latitudinal center $\lambda$ at 13$^\circ$.43 of the northern hemisphere. In contrast to AR 12673, its tilt angle has small variations. The $\sigma_\alpha$ during September 1st to 9th, when its LDCM was less than 60$^\circ$, is 2$^\circ$.87. The average tilt $\overline{\alpha}$ is 22$^\circ$.2. The main information on the two ARs is listed in Table \ref{tab:ARs}.

\begin{deluxetable*}{cccccccccccccc}
\tablenum{1}
\tablecaption{Physical quantities of ARs 12673 and 12674 discussed in the paper \label{tab:ARs}}
\tablewidth{0pt}
\tablehead{
\colhead{AR No.} & \colhead{TYPE} & \colhead{N/a-N} & \colhead{$A$} &
\colhead{$\lambda$} & \colhead{$\overline{\alpha}$} & \colhead{$\sigma_\alpha$} & \colhead{$F_{+}$} & \colhead{$F_{-}$} & \colhead{$F$} & \colhead{$|B_p|$} & \colhead{$|D|$} & \colhead{$DI$} & \colhead{$FI$}
}
\startdata
12673 &$\beta\gamma\delta$ & a-N&  1397   & -9.36 & $7^{\circ}.0$ & $43^{\circ}.16$ & 1.74 & -1.85 & 3.50  & 0.06 & 0.034 & 27.34 & 2978.8 \\
12674 &$\beta$ & N & 1238   & 13.43 & $22^{\circ}.2$ & $2^{\circ}.87$ & 1.18 & -1.60 & 3.13 & 0.08 & 0.044 & 33.47 & 36.6 \\
\enddata
\tablecomments{`TYPE' is the Mount Wilson sunspot classification provided by NOAA/USAF. `N/a-N' denotes the normal/ abnormal polarities of ARs. $A$ is the maximum area in $\mu$Hem. $\lambda$ is the corresponding latitudinal location. $\overline{\alpha}$ and $\sigma_\alpha$ are the mean tilt angle and the standard deviation, respectively, when ARs are within 60$^\circ$ LDCM. $F_{+}$ and $F_{-}$ are the positive flux and negative flux in $10^{22}$ Mx, respectively, based on the HMI CR2195 synoptic map. $F$ is the corrected total flux of each AR with balanced positive and negative fluxes. $B_p$ and $D$ are the absolute values of the final polar field and axial dipole moment in Gauss generated by each AR in SFT simulations. $DI$ and $FI$ are the dipole moment index and flare index defined in the paper.}
\end{deluxetable*}


\subsection{Their effects on space weather}
The Geostationary Operational Environmental Satellite (GOES) data show that AR 12673 produced 27 M-class and 4 X-class flares (X9.3, X8.2, X2.2, and X1.3). However, AR 12674 did not produce any flares over the M1.0 class. The strongest flare produced by this AR is C5.2. The flare index ($FI$) was first introduced by \cite{Antalova1996} and was later widely applied by other authors to quantify the flare productivity of a given AR \citep[e.g.,][]{Abramenko2005,Romano2007,Chen2011}. We follow the traditional definition by weighting the soft X-ray flares of classes C, M, and X by 1, 10, and 100, respectively, regardless of the duration of the flare. To be more explicit, we provide it in the following format:
\begin{equation}
\label{eq:fi}
FI=\sum M_C+10\times\sum M_M+100\times\sum M_X,
\end{equation}
where $M_C$, $M_M$, and $M_X$ are the magnitudes of all flares belonging to the classes C, M, and X, respectively. Flares smaller than the C1.0 class are not considered in the definition. The $FI$ values of ARs 12673 and 12674 are 2978.8 and 36.6, respectively. The values clearly demonstrate that AR 12673 had stronger effects on space weather than AR 12674. In the following subsection, we compare their effects on space climate by investigating their contributions to the solar cycle variation.

\begin{figure}[htp]
\begin{center}
\includegraphics[scale=0.4]{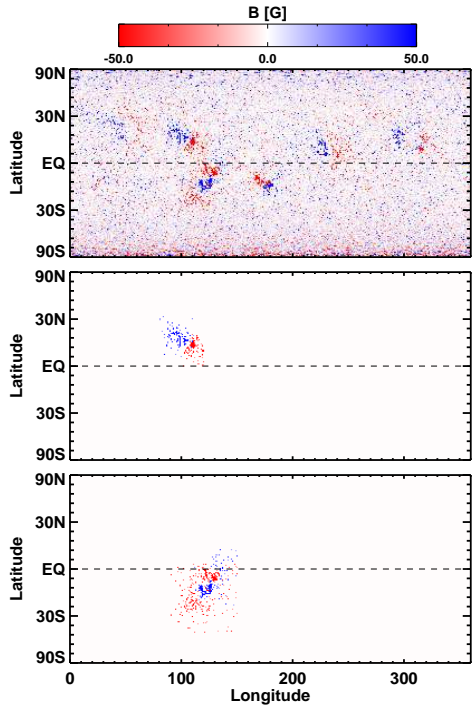}
\caption{Synoptic map of the photospheric radial magnetic field of CR2195 from the Helioseismic and Magnetic Imager (HMI) on board the Solar Dynamics Observatory (SDO) spacecraft (upper panel) and the isolated active regions 12683 (middle panel) and 12682 (lower panel). ARs 12682 and 12683 are the recurrent ARs 12673 and 12674 after one CR, respectively.} \label{fig:MagARs9394}
\end{center}
\end{figure}

\subsection{Their effects on space climate}
Here we investigate different contributions from different ARs to the long-term evolution of the polar field and axial dipole moment, which reflect their effects on space climate. The upper panel of Figure \ref{fig:MagARs9394} is the HMI/SDO Synoptic Chart of the surface radial magnetic field of Carrington Rotation (CR) 2195 (2017 September 12th - October 10th). A calibration factor to MDI/SOHO of 1.3 is adopted here \citep{Liu2012}. ARs 12673 and 12674 reappeared around the central meridian on October 1st and October 2nd, respectively, one CR later after their first appearance on the solar disk. They were denominated as ARs 12682 and 12683, respectively.

SFT simulations can reproduce well the observed large-scale solar surface field  \citep[e.g.][]{Wang1989b,Baumann2004,Jiang2010,Mackay2012, Jiang2014a, Upton2014}. The emergent magnetic flux corresponding to the sunspot groups at the solar surface provides the source of the surface flux. The flux is then transported and dispersed over the solar surface by systematic flows, including meridional flow and differential rotation, and turbulent motions that are effectively treated as turbulent diffusion. These processes are modeled by the surface flux transport equation. The equation is the $r$-component of the MHD induction equation at the solar surface under the assumption that the field at the surface is purely vertical.

Here we adopt the SFT model that has been used in our recent studies, i.e., \cite{Jiang2015, Jiang2018b}. All of the transport parameters are constrained from the available observations. The only difference between the current model and previous models \citep{Jiang2015, Jiang2018b} is the strategy of incorporating the source term of flux emergence. In the past simulations, we used the analytical bipolar magnetic flux distributions for all the ARs or assimilated the full synoptic magnetograms as the initial condition in the solar cycle predictions \citep{Cameron2016, Jiang2018a, Jiang2018b}. Apparently, these methods are not realistic for isolating the contribution from a complex multipolar configurations, like $\delta$-type ARs. We assimilate each observed AR that is isolated from the observed synoptic map into our SFT model following the idea of \citet{Yeates2015} and \citet{Whitbread2018}, who have inserted the observed shapes of individual ARs in their SFT simulations.
The lower two panels of Figure \ref{fig:MagARs9394} are the isolated two ARs 12682 (12673) and 12683 (12674). The timing of the magnetic flux distribution assimilated into the SFT model is a major difference from the assimilation technique by \cite{Yeates2015} and \cite{Whitbread2018}. They assimilated ARs into their simulations on the days of central meridian crossing. Big ARs usually have not reached their area maximum on the day of the first central meridian crossing, and magnetic configurations usually change remarkably during flux emergence. So, we take the time around their second central meridian crossing to start the simulations. This corresponds to ARs 12682 and 12683 on the CR2195 synoptic map. After the second central meridian crossing of an AR, there is usually no further flux emergence, and the configuration becomes stable. This makes it a reasonable time for the SFT simulation to obtain the contribution of the individual AR to the axial dipole moment. Although a certain amount of flux might have been cancelled before the assimilation into the SFT simulation, it does not affect its contribution to the final axial dipole moment, which corresponds to the large-scale and slow-decay field.

The major problem we encounter while assimilating the observed individual ARs into the SFT model is the unbalanced magnetic flux. The positive and negative fluxes for AR 12683 (12674) are 1.18$\times10^{22}$ Mx and -1.60$\times10^{22}$ Mx. We assume that some positive flux is too diffusive to get identified. Hence, we increase each positive pixel by the same percentage as the difference between the two polarities, i.e., 35.6\%. This is another difference of our assimilation method from the assimilation method used by \cite{Yeates2015}. Then, we extrapolate the equal sine latitude distribution of the HMI synoptic magnetogram to equal latitude. We used the IDL Congrid program to reduce the size of the entire domain to 360 pixels in longitude and 180 pixels in latitude, which is the resolution of our SFT simulations. After all these procedures, the total flux assimilated into the SFT model is 3.13$\times10^{22}$ Mx with balanced positive and negative fluxes. The positive and negative fluxes for AR 12682 (12673) are  1.74$\times10^{22}$ Mx and -1.85$\times10^{22}$ Mx, respectively, with a slight imbalance. The same procedure is followed to assimilate AR 12682 (12673). The final total flux assimilated to the SFT simulation for AR 12682 (12673) is 3.50$\times10^{22}$ Mx.

\begin{figure}[htp]
\plottwo{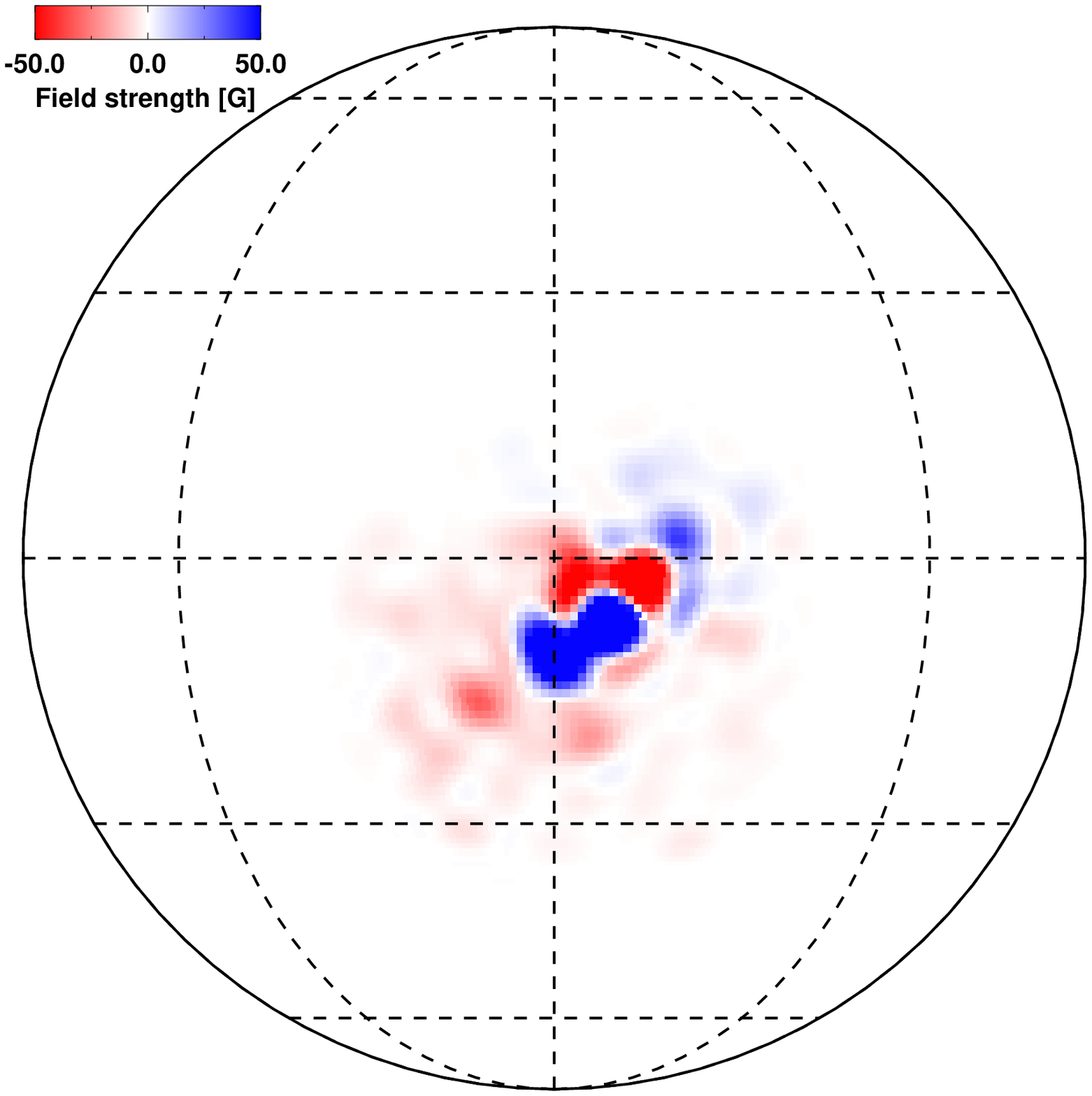}{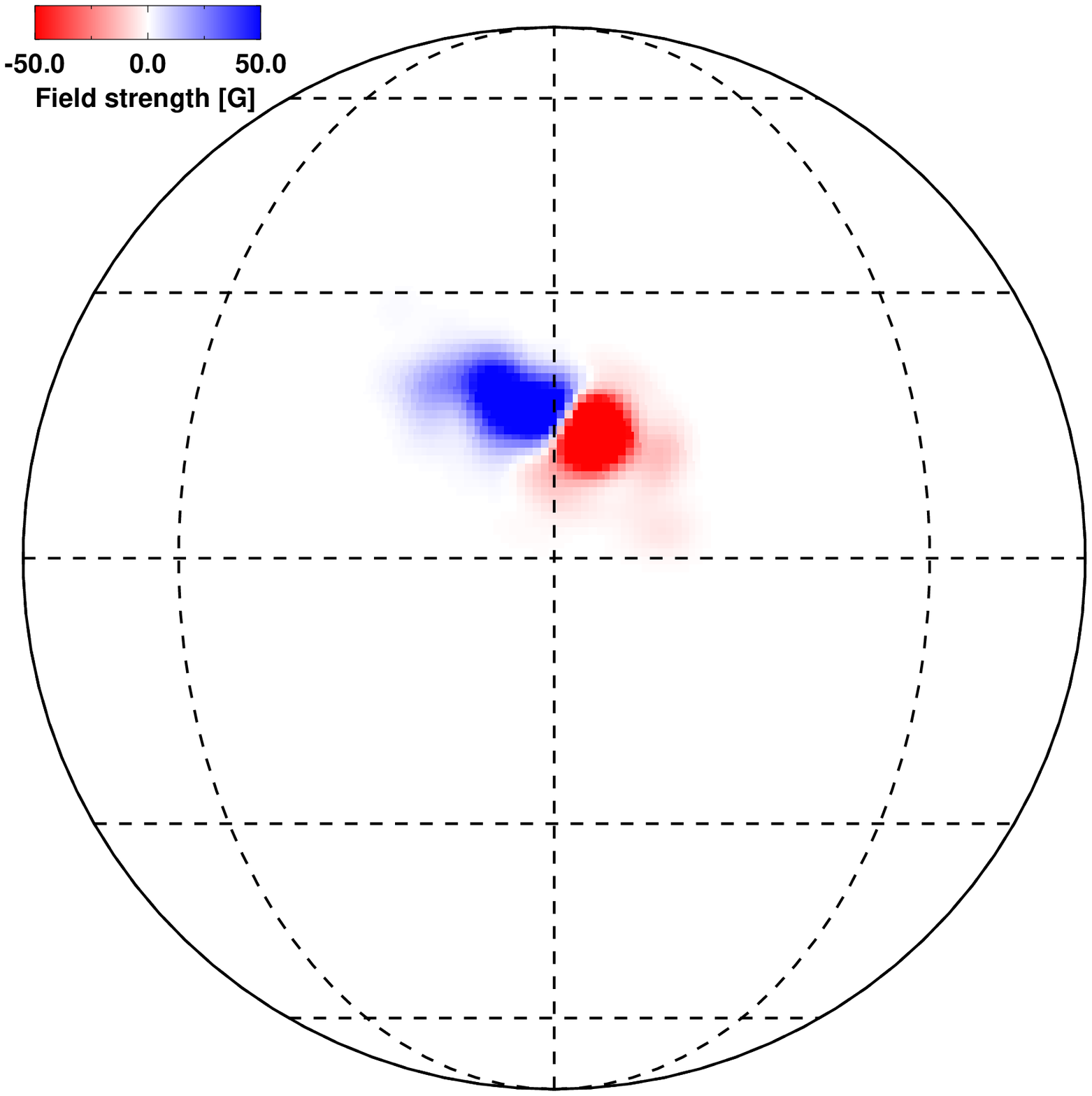}
\caption{Orthographic projections of AR 12682 (the recurrent AR 12673, left panel) and AR 12683 (the recurrent AR 12674, right panel) centered at the central meridian and the equator. They correspond to decompositions of the observed magnetograms shown in the lower two panels of Figure \ref{fig:MagARs9394} in spherical harmonics. The animation shows the time evolution of ARs 12673 and 12674. The videos start at 2018 October 1 for AR 12673 and 2018 October 2 for AR 12674. They persist for 10 years with an interval of 27 days. \label{fig:br_prjt}}
\end{figure}

We take the observed synoptic maps shown in the lower two panels of Figure \ref{fig:MagARs9394} as the initial conditions of the SFT simulations for AR 12674 and AR 12673. The flux transport equation on the surface of a sphere is solved by  decomposing the radial magnetic field into spherical harmonics, originally developed by \cite{Baumann2004}. We take the maximum order of the spherical harmonics $l=64$. The discontinuous discrete data from observations naturally bring spectral leakage. A Hann window is used to reduce the side lobes (ripple) of the harmonics response. Figure \ref{fig:br_prjt} shows orthographic projections of AR 12682 (12673) and AR 12683 (12674) centered at the central meridian and the equator after using the spherical harmonics decomposition of the isolated ARs presented in lower two panels of Figure \ref{fig:MagARs9394}. We run each simulation for 10 years to make sure that the polar field reaches a final balanced state \citep{Ballegooijen1998}. Animations of the time evolution of the two ARs starting from the time of data assimilation with a time interval of 27 days are available.

\begin{figure}
\begin{center}
\includegraphics[scale=0.4]{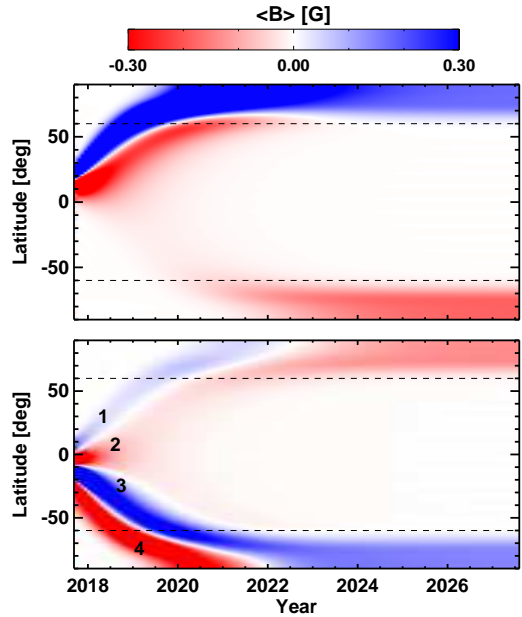}
\caption{Time evolution of the longitudinally averaged magnetograms derived based on the SFT simulations of the individual AR 12674 (upper panel) and AR 12673 (lower panel).} \label{fig:MagBftARs9394}
\end{center}
\end{figure}

\begin{figure}[htp]
\begin{center}
\includegraphics[scale=0.35]{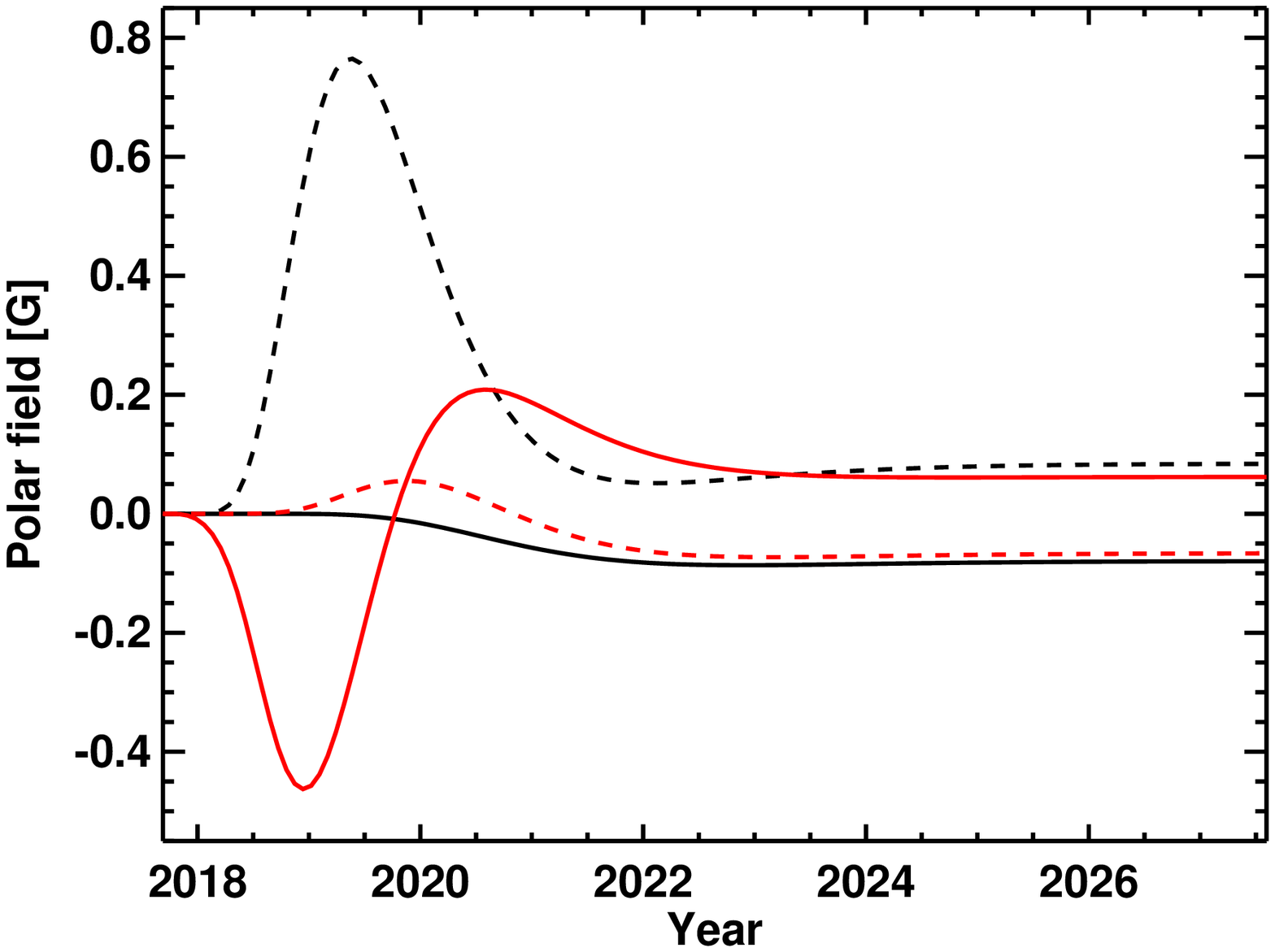}
\includegraphics[scale=0.35]{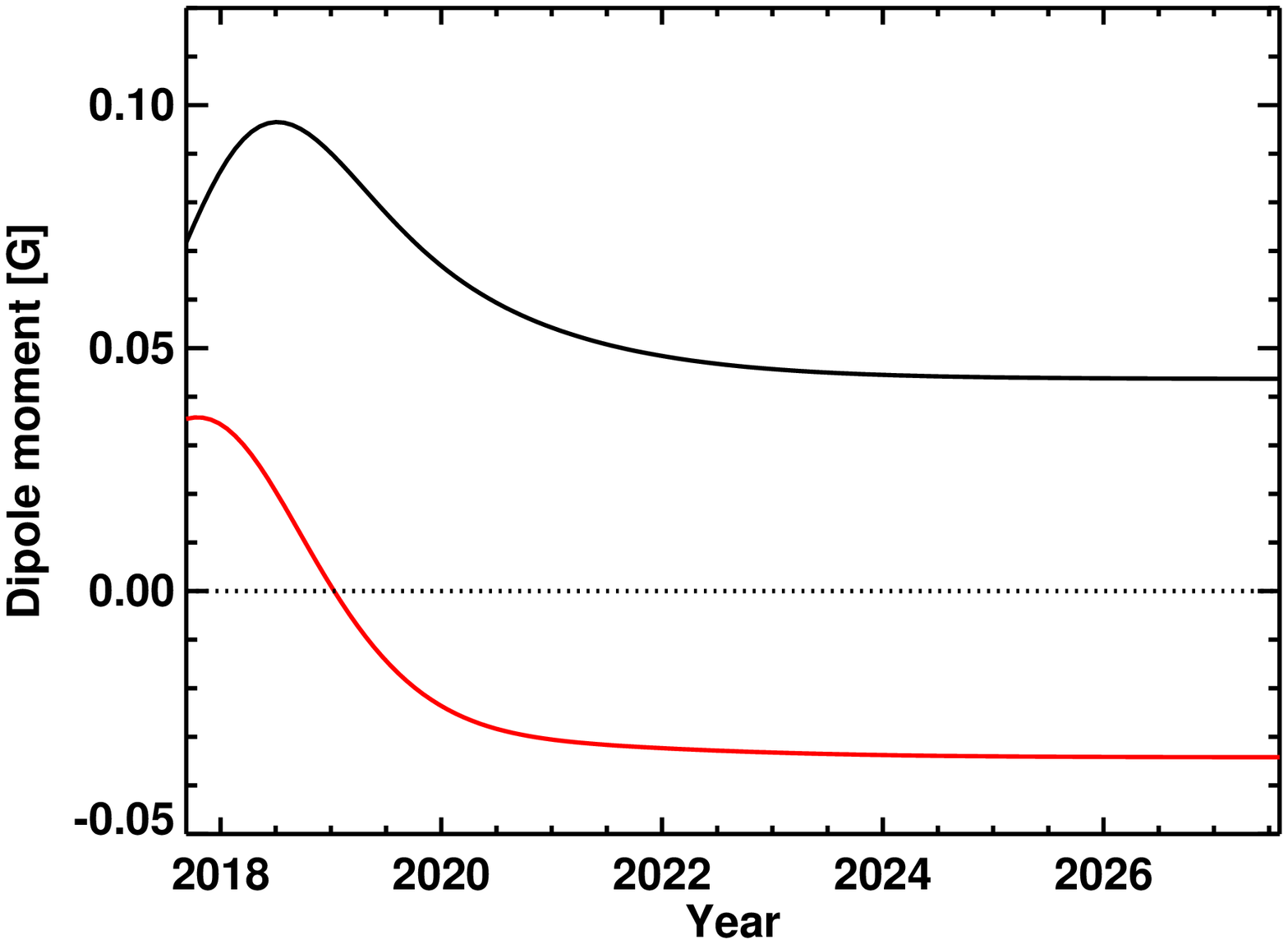}
\caption{Time evolution of the polar field (left panel) and the axial dipole moment (right panel) from the SFT simulations of AR 12673 (red curves) and AR 12674 (black curves). The north polar field is shown by the dashed curves and the south polar field by the solid curves.} \label{fig:PF_DM}
\end{center}
\end{figure}

Figure \ref{fig:MagBftARs9394} shows the time evolution of the longitudinally averaged magnetic field. The upper panel is for the simple configuration of AR 12674. It shows a typical evolution of an AR using SFT simulations presented in the previous studies \cite[e.g.,][]{Wang2000,Mackay2002,Jiang2010,Yeates2015}. As AR 12674 is bipolar, part of the leading negative polarity diffuses across the equator. Under the effects of the poleward meridional flow and the turbulent diffusion, the negative flux that crossed the equator is finally transported to the southern pole. The corresponding positive flux is transported to the northern pole. Finally, a north-south balanced polar field sets up after about 6 years. Equations (4) and (5) of \cite{Jiang2018a} are used to calculate the polar field $B_p(t)$ that is averaged over the $\pm75^{\circ}$ to $\pm60^{\circ}$ latitudes in each hemisphere and the axial dipole moment $D(t)$, respectively. The black curves in Figure \ref{fig:PF_DM} show the time evolution of $B_p$ and $D$. The arrival of the following positive polarity increases the northern polar field. The arrival of the leading negative polarity that cancels with the existing positive flux causes the later decrease of the northern polar field. The negative leading flux gets transported to the southern pole and generates the negative polar field there. The final balanced northern and southern polar fields becomes 0.08 and -0.08 G, respectively. The whole process corresponds, first, to an increase of the axial dipole moment due to the separation of the two polarities and then to a decrease due to the flux cancellation. The final axial dipole moment $D_f$ is 0.044 G. According to Figure 8 of \cite{Jiang2018b}, the axial dipole moments generated by all ARs during cycles 21-23 are 8.2 G, 7.4 G, and 4.3 G, respectively. On average, the axial dipole moment generated by all of the ARs during a cycle is 6.6 G. This means that about 150 ARs of such kind, which correspond to less than 7\% of all ARs are enough to be responsible for the solar cycle evolution. To obtain this estimation, we have assumed that all ARs have normal polarities.

The lower panel of Figure \ref{fig:MagBftARs9394} represents the time evolution of the longitudinally averaged magnetic field for the complex AR 12673. The red curves in Figure \ref{fig:PF_DM} are the time evolution of $B_p$ and $D$. The results are remarkably different from the simple case. It seems like two bipolar structures contributing to $B_p$ and $D$. The bipolar polarities denoted as `1' and `2' emerge across the equator. The negative polarity has a much stronger flux, which produces the strong cross-equator negative flux. The positive flux arrives at the polar region first and hence generates a positive polar field. The negative polarity arrives at the polar region later. The negative polarity first cancels the positive flux and then builds up the negative flux in the northern pole.  The other bipolar polarities, denoted as `3' and `4' have stronger positive flux. The lower latitude negative flux arrives at the southern pole first and forms the negative polar field, and then it gets reversed by the stronger positive flux. Hence, we see the reversals of the polar field indicated by the red curves in the left panel of Figure \ref{fig:PF_DM}. The final balanced northern and southern polar fields are -0.06 and 0.06 G, respectively. The resultant polarity it generates is opposite to that generated by AR 12674, making the polar field weaker at cycle minimum. This indicates that the AR 12673 has abnormal polarity.

The axial dipole moment evolution of AR 12673 shows an interesting result, which seems surprising and has not been reported before to our knowledge. Its initial axial dipole moment $D_0$ is positive, which is consistent with the cycle. However, the final axial dipole moment is opposite to the initial one. The value of $D_0$ is proportional to $B\sin\lambda\cos\lambda$. Hence, the flux closer to the equator, e.g., the polarities denoted as `1' and `2' in Figure \ref{fig:MagBftARs9394} has less contribution to $D_0$ than the flux from higher latitudes, e.g., the polarities denoted as `3' and `4'. The polarities `3' and `4' dominate the positive $D_0$ value, but the final axial dipole moment from the AR is determined by the cross-equator flux, which is dominated by the cross-equator polarity `2'. Most of its negative flux is transported to the northern pole. Eventually, a negative/positive polar field is built up in the northern/southern poles. This corresponds to a strongly negative axial dipole moment. In short, it is the flux transport and the initial magnetic configuration that cause the reversal of the axial dipole moment and significantly weaken the axial dipole moment at the end of cycle 24. This numerical experiment demonstrates the importance of including the configuration of the $\delta$-type ARs into SFT models to study their contributions to the large-scale field. The widely adopted methods, which use the tilt angle to simplify the initial configuration, cannot get the reversal of the axial dipole moment presented here.

Here, we only considered the long-term (or the final) contribution to the polar field and the axial dipole moment from the individual ARs. The significant transient perturbations to the polar field due to strong pairs of opposite plumes in both panels of Figure \ref{fig:MagBftARs9394} are not what we aim to address here because their effects on the solar cycle have not been clear until now.

\begin{figure}[htp]
\begin{center}
\includegraphics[scale=0.4]{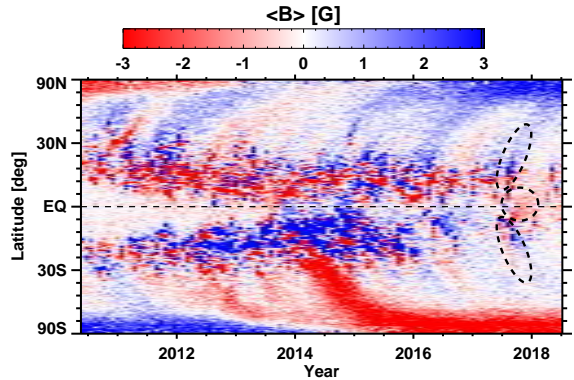}
\caption{Time evolution of the longitudinal averaged synoptic maps from SDO/HMI. The three typical structures produced by ARs 12673 and 12674 are denoted by the overlying ellipses.} \label{fig:MagBftHMI}
\end{center}
\end{figure}

The time evolution of the longitudinally averaged synoptic magnetograms observed by SDO/HMI is shown in Figure \ref{fig:MagBftHMI}. From the annotated ellipses, we can see the following structures that are consistent with the simulated results. First, there is a positive poleward plume followed by the relatively weaker negative plume produced by AR 12674 in the northern hemisphere. Second, there is more than one-year persistence of the negative flux around the equator produced by the cross-equator emergence of the negative flux of AR 12673 and migration of the leading polarity of AR 12674 towards the southern pole. Third, there is a weak negative poleward plume followed by the stronger positive poleward plume in the southern hemisphere produced by AR 12673. Please note that remnants of other ARs make the observed structures not that clear.

We have shown that the big $\beta\gamma\delta$-type AR 12673 has strong effects on both the short-term and the long-term solar activity. Although the big $\beta$-type AR 12674 does not generate strong space weather events, it has a large contribution to the dipole moment at the end of the cycle, and hence on space climate. In the following section, we perform a statistical study of big solar ARs to investigate their effects on the short-term and long-term variations of solar activity.


\section{Statistical studies of big solar active regions}
\label{sec:Statistics}

\subsection{Data source}
\label{sec:data}
The DPD is the most detailed and user-friendly catalog at present \citep{Baranyi2016}. It provides the area and position data for each observable sunspot and sunspot groups on a daily basis along with images of sunspot groups, full-disk scans, and magnetograms starting from 1974 to the present \citep{Baranyi2001,Baranyi2013}. The data are stable over the available time period, and the calibration factor to the RGO dataset is close to 1.0 \citep{Gyori2017}.
The tilt angle data of the sunspot groups are also available from the DPD. The tilt angles of the sunspot groups are determined in a similar way to the longest datasets of sunspot group tilt angles from Mount Wilson Observatory and Kodaikanal Solar Observatory  \citep{Baranyi2015}. \cite{Baranyi2015} indicated that the tilt angle data do not contain any magnetic polarity information on spots, but the available magnetograms are frequently taken into account while grouping spots. Thus, the DPD tilt angles data based on estimated polarities are close to the data where magnetic polarities are taken into account \citep[e.g.][]{Li2012,Stenflo2012}. Therefore, we have considered the area, position, and tilt angle of each AR based on the DPD data set.

For each AR, its area varies when it passes across the solar disk. We designate its area $A$ as its maximum area when it is within 60$^\circ$ LDCM. The corresponding location is designated as the AR's location, including the latitude $\lambda$. The tilt angle of each AR also varies during its passing across the solar disk. We take the averaged value of the tilts, $\overline{\alpha}$, when it is within 60$^\circ$ LDCM as the tilt angle of the AR. The corresponding standard deviation of the tilt angle is $\sigma_\alpha$.

The GOES X-ray data are available from 1975 September onwards\footnote{https://www.ngdc.noaa.gov/stp/space-weather/solar-data/solar-features/solar-flares/x-rays/goes/xrs/}. So, we select the ARs that are larger than 800 $\mu$Hem from the beginning of 1976 to the end of 2017. The 567 ARs that satisfy this requirement are singled out as our samples. The samples show that 30\% of ARs with areas larger than 1000 $\mu$Hem are able to produce flares over X1. We use Equation (\ref{eq:fi}) to calculate each sample's flare index $FI$.

The polar field at solar cycle minimum, which is the source of the toroidal flux that emerges in the subsequent cycle under the framework of the BL-type dynamo, is generated by the emergence and evolution of tilted sunspot groups. The latitudinal separation and areas of the positive and negative flux dominate the initial axial dipole moment of the emerged AR. The subsequent flux transport processes over the surface depend on the latitudinal distribution of the magnetic flux and transport parameters, which introduce big differences between the initial and the final axial dipole moments that an AR contributes to the solar cycle. \cite{Jiang2014b} gave an empirical relation of the latitudinal dependence of the final axial dipole moment, which is in the form of $\exp{(-\lambda^2/110.0)}$. Other studies show a slight difference in the HWHM of the Gaussian function \citep{Nagy2017,Whitbread2018}. This is due to the meridional flow speeds assumed at low latitudes, which have effects on the rate and amount of cross-equatorial diffusion. Based on the dependence of the final axial dipole moment on the area, tilt angle, and latitudinal location, we define a dipole moment index $DI$ as the proxy of the ARs' effects on the variation of the solar cycle and on space climate. The form is as follows:
\begin{equation}
\label{eq:di}
DI=A_s\sin{|\overline{\alpha}|}\exp{(-\lambda^2/110.0)},
\end{equation}
where $A_s=A+A_f$ is in units of degree square. The facular area $A_f$ is connected to the sunspot area $A$ by $A_f=414+21A-0.0036A^2$ in units of $\mu$Hem. The method is less realistic than the method used in Section \ref{sec:2ARs} to calculate the final axial dipole moment, especially for the $\delta$-type ARs. But the proxy provides a quick and convenient estimation of the contribution to space climate. The $DI$ values for ARs 12673 and 12674 are 27.34 and 33.47, respectively. Their relative amplitudes are consistent with the detailed calculations in Section \ref{sec:2ARs} although the estimation cannot distinguish the positive and the negative contributions to the axial dipole moment. Moreover, for the statistical analysis here, we do not distinguish the positive and negative contributions to the solar cycle variation.

The magnetic complexity of ARs has a close correlation with flare eruptions. Here, we use Mount Wilson magnetic classifications of ARs to indicate the complexity. There are two datasets for the classifications. One is from the Mount Wilson sunspot record \footnote{ftp://ftp.ngdc.noaa.gov/STP/SOLAR\_DATA/SUNSPOT\_REGIONS/Mt\_Wilson/} for ARs during 1976-2004 and the other dataset is the NOAA/USAF/SOON dataset \footnote{http://solarcyclescience.com/activeregions.html} for ARs during 2005-2017. The NOAA numbering system of ARs was not used in the early time of the dataset. We match the ARs to get their classifications using the following method. If there are observations on the day of a given AR at its maximum area, we match the ARs by comparing the latitude difference (within 2$^\circ$) and the difference in LDCM (within 10$^\circ$). If there are no observations on the day of a given AR, the data on the nearby date are used with the consideration of the shift of the LDCM. Due to the evolution of ARs, a given AR has different types on different days when it passes across the solar disk. When the $\delta$ structure appears more than twice, the AR is designated as a $\delta$-type AR. For simplicity, all of the ARs are classified into two groups, that is, a complex type that includes the $\delta$ structure (i.e., $\beta\gamma\delta$ or $\gamma\delta$) and a simple type that does not include the $\delta$ structure (i.e., $\alpha$, $\beta$, $\gamma$, or $\beta\gamma$).

Table \ref{tab:listARs} gives the detailed information on the 567 big ARs including the time, NOAA No., latitude, area, magnetic classification, mean tilt angle, standard deviation of tilt angles, axial dipole moment index, flare index, and maximum fare class.

\begin{deluxetable*}{ccccccccccc}
\tablenum{2}
\tablecaption{Detailed information on the ARs larger than 800$\mu$Hem from 1976 to 2017 \label{tab:listARs}}
\tablewidth{0pt}
\tablehead{
\colhead{No.} & \colhead{Time} & \colhead{AR No.} & \colhead{$\lambda$} &
\colhead{$A$} & \colhead{TYPE} & \colhead{$\overline{\alpha}$} & \colhead{$\sigma_\alpha$} & \colhead{$DI$} & \colhead{$FI$} & \colhead{$Fm$}
}
\startdata
  1&  19760328  &   690  &   -7.64  &   919  &    D     &    8.34   &   20.53   &   30.96   &  328.0   &  100.0  \\
  2&  19770626  &   839  &   14.56  &  1269  &    D     &   21.16   &   15.10   &   24.42   &   69.0   &   20.0  \\
  3&  19770912  &   889  &    7.90  &  1086  &    D     &    8.06   &   14.27   &   32.90   &  708.0   &  200.0  \\
  4&  19780103  &   969  &   20.77  &   928  &    B     &    1.84   &    4.83   &    0.23   &   22.0   &    7.0  \\
  5&  19780213  &  1001  &   15.19  &  2286  &    B     &   16.00   &    1.78   &   22.26   &  390.0   &   70.0  \\
\enddata
\tablecomments{The table includes the time in `yyyymmdd', NOAA AR No., latitude $\lambda$, maximum area $A$, magnetic classification `TYPE' (`D' corresponds to the complex type, `B' corresponds to the simple type), mean tilt angle $\overline{\alpha}$, standard deviation of tilt angles $\sigma_\alpha$, axial dipole moment index $DI$, flare index $FI$ and maximum flare class $Fm$ of 567 ARs. This table is available in its entirety in a machine-readable form in the online
journal. A portion is shown here for guidance regarding its form and content.}
\end{deluxetable*}

\subsection{Statistical results}
\label{sec:results}

Among the 567 big ARs, the percentages of simple and complex ones are 42\% and 58\%, respectively. The average areas of the complex ARs and simple ARs are 1453 $\mu$Hem and 1116 $\mu$Hem, respectively. These show a weak trend for big ARs to be in the complex configurations.

\begin{figure}[htp]
\begin{center}
\includegraphics[scale=0.4]{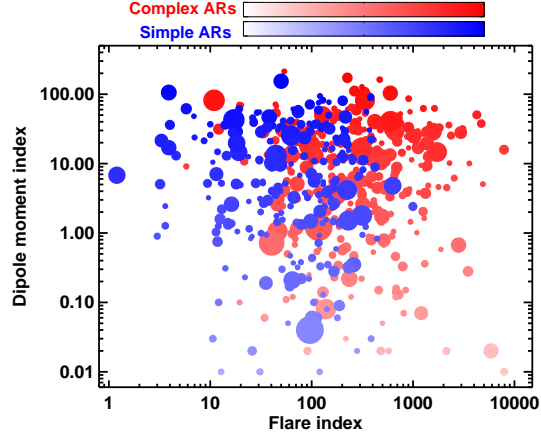}
\caption{Scatter plot of the flare index $FI$ and axial dipole moment index $DI$ for the 567 selected AR samples. The size of each point corresponds to the area of each AR. Simple ARs are shown in blue, and complex ARs are shown in red. The color opacity corresponds to the amplitude of $\sin{|\overline{\alpha}|}\exp{(-\lambda^2/110.0)}$.} \label{fig:FIvsDM}
\end{center}
\end{figure}

Figure \ref{fig:FIvsDM} shows the scatter plot of the axial dipole moment index $DI$ and the flare index $FI$ for all of the big ARs. The complex ARs in red dots tend to have strong space weather effects. The percentage of complex ARs with $FI$ larger than 100, which is equivalent to the flare class X1, is 81.1\%. The weak space weather effects of the rest of the complex ARs mainly result from our simplification of the definition of the AR magnetic classification. We designate an AR as complex when the $\delta$ configuration is recorded two or more times when it is within 60$^\circ$ LDCM. Over a majority of an AR's lifetime, the AR probably is in the $\beta$ structure. The percentage of simple ARs that have $FI$ smaller than 200, which is equivalent to an X2 flare, is 85.0\%. Simple ARs that have larger FIs usually include mixed polarities, which correspond to the relative complex $\gamma$ structure. The axial dipole moment index $DI$ has a strong dependence on the latitude and the tilt angle of the AR. The red/blue opacity corresponds to the amplitude of $\sin{|\overline{\alpha}|}\exp{(-\lambda^2/110.0)}$. Simple and complex ARs have no difference in latitudinal emergence. The mean absolute value of the tilt angle of both types of ARs is about 10$^\circ$.5, which is much larger than the mean tilt angle of ARs, 5$^\circ$.29 \citep{Baranyi2015}. Furthermore, complex ARs usually have large variations of tilt angles, which cause a large uncertainty for the estimation of $DI$. The mean standard deviations of the tilt angle of simple and complex ARs are 10$^\circ$.9 and 12$^\circ$.2 respectively. The percentages of simple and complex ARs whose $DI$ are larger than 10 are 42.7\% and 49.7\%, respectively. Overall, simple ARs have the same impact as complex ARs on the solar cycle variation, and hence on space climate.

\begin{figure}[htp]
\begin{center}
\includegraphics[scale=0.4]{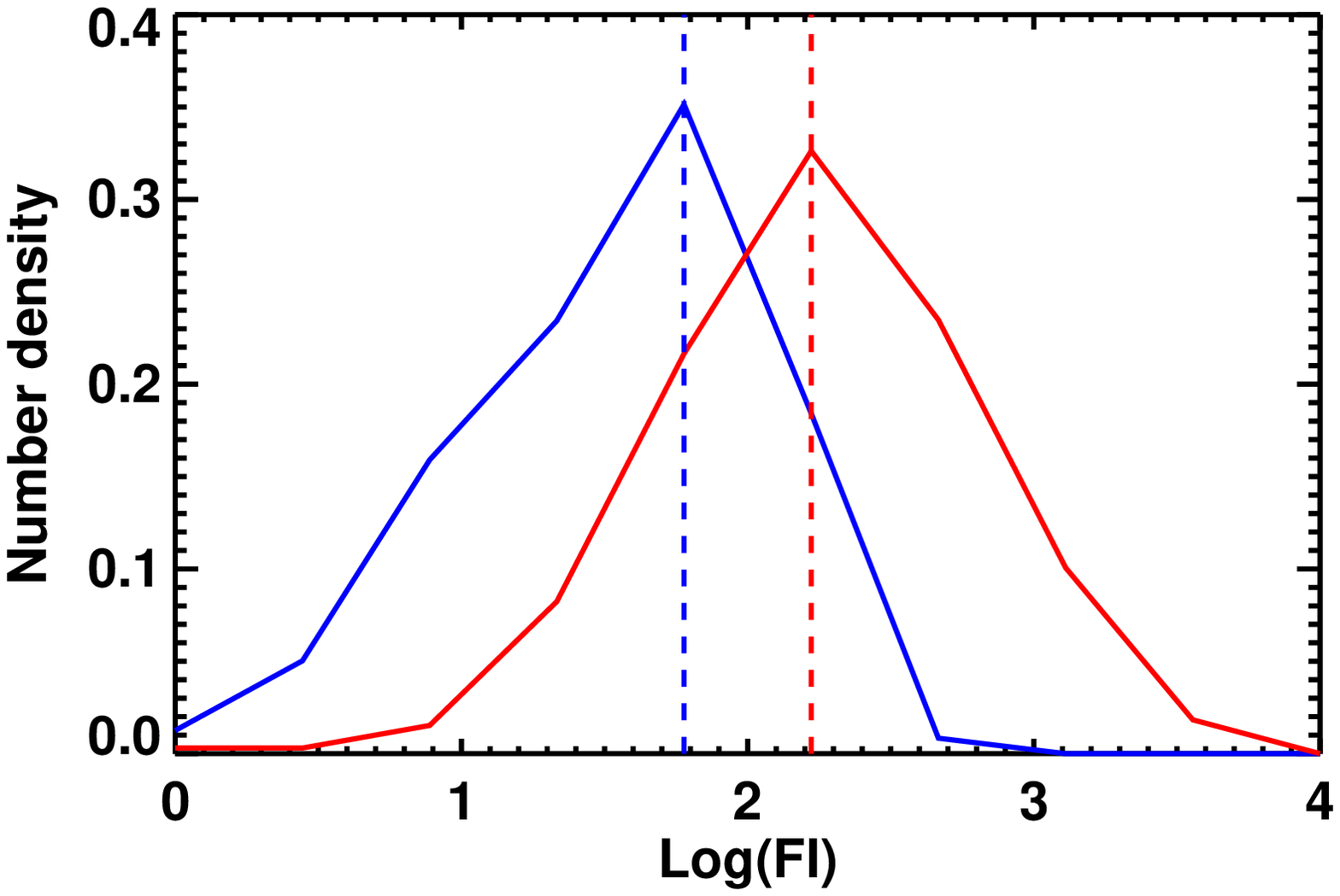}
\includegraphics[scale=0.4]{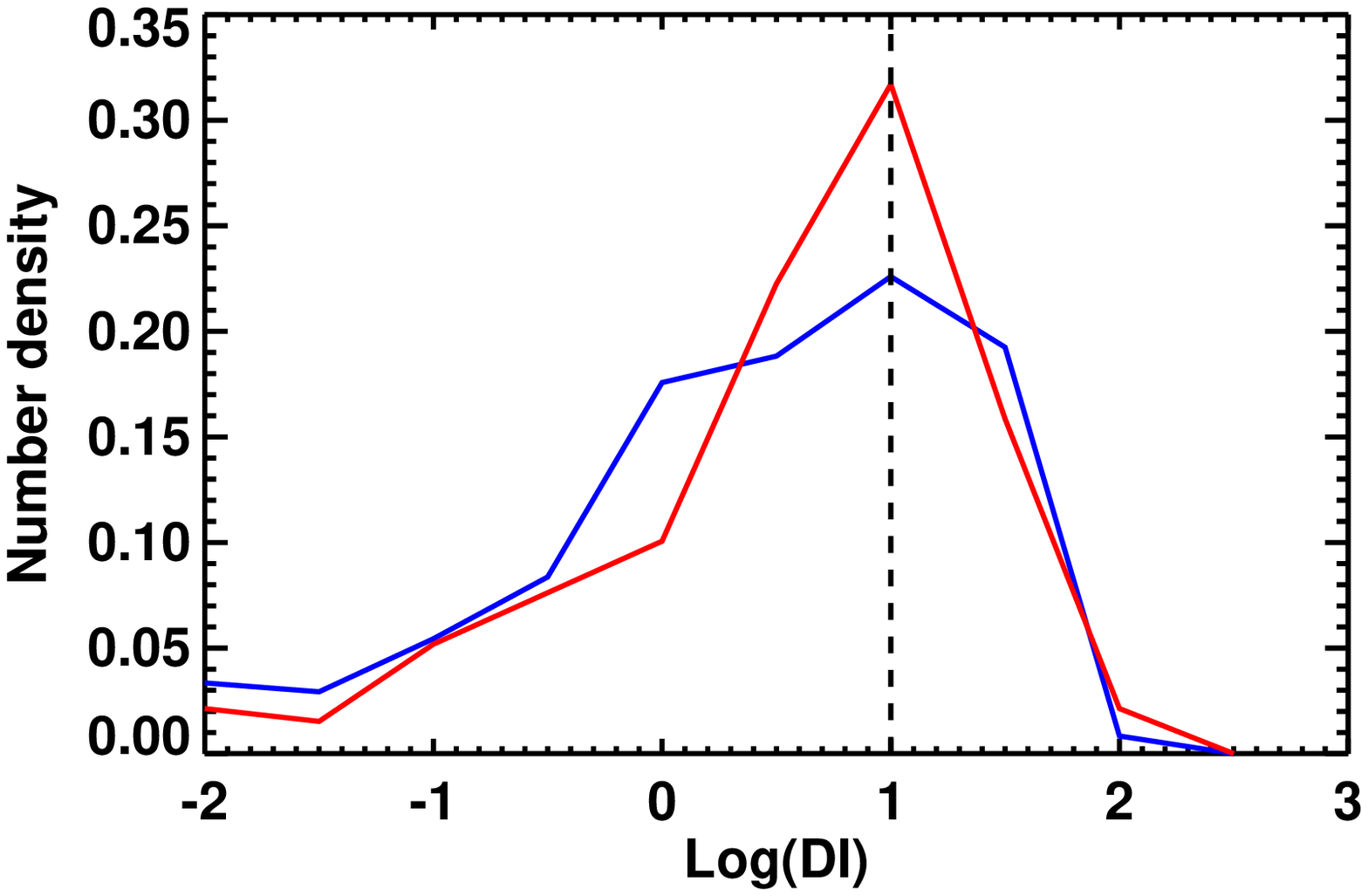}
\caption{Normal number density distribution of the logarithm values of the flare index $FI$ (left panel) and the axial dipole moment index $DI$ (right panel). The results for the simple ARs are shown in blue, and those for the complex ARs are shown in red. The vertical lines are the locations of the maximum number density distribution.} \label{fig:FI_DM_pdf}
\end{center}
\end{figure}

Figure \ref{fig:FI_DM_pdf} shows the normal number density distribution of the logarithm values of the flare index (left panel) and the axial dipole moment index (right panel). For the flare index, the number density distributions for both simple and complex ARs have similar profiles, but different maximum number density location. For the simple ARs, the flare index $FI$ is mainly concentrated in the bin ranging from 60.2 ($\log FI=1.78$) to 166.0 ($\log FI=2.22$). For the complex ARs, the flare index $FI$ is mainly concentrated in the bin ranging from 166.0 ($\log FI=2.22$) to 457.1 ($\log FI=2.66$). By contrast, the dipole moment index of the simple and complex ARs have the same maximum number density locations. They both correspond to the bin of $DI$ values in the range of 10 ($\log DI=1.0$) to 31.6 ($\log DI=1.5$). The complex ARs tend to be concentrated in the bin of maximum number density. The simple ARs tend to have a uniform distribution in their $DI$ values from 1 to 100.

\begin{figure}[htp]
\begin{center}
\includegraphics[scale=0.4]{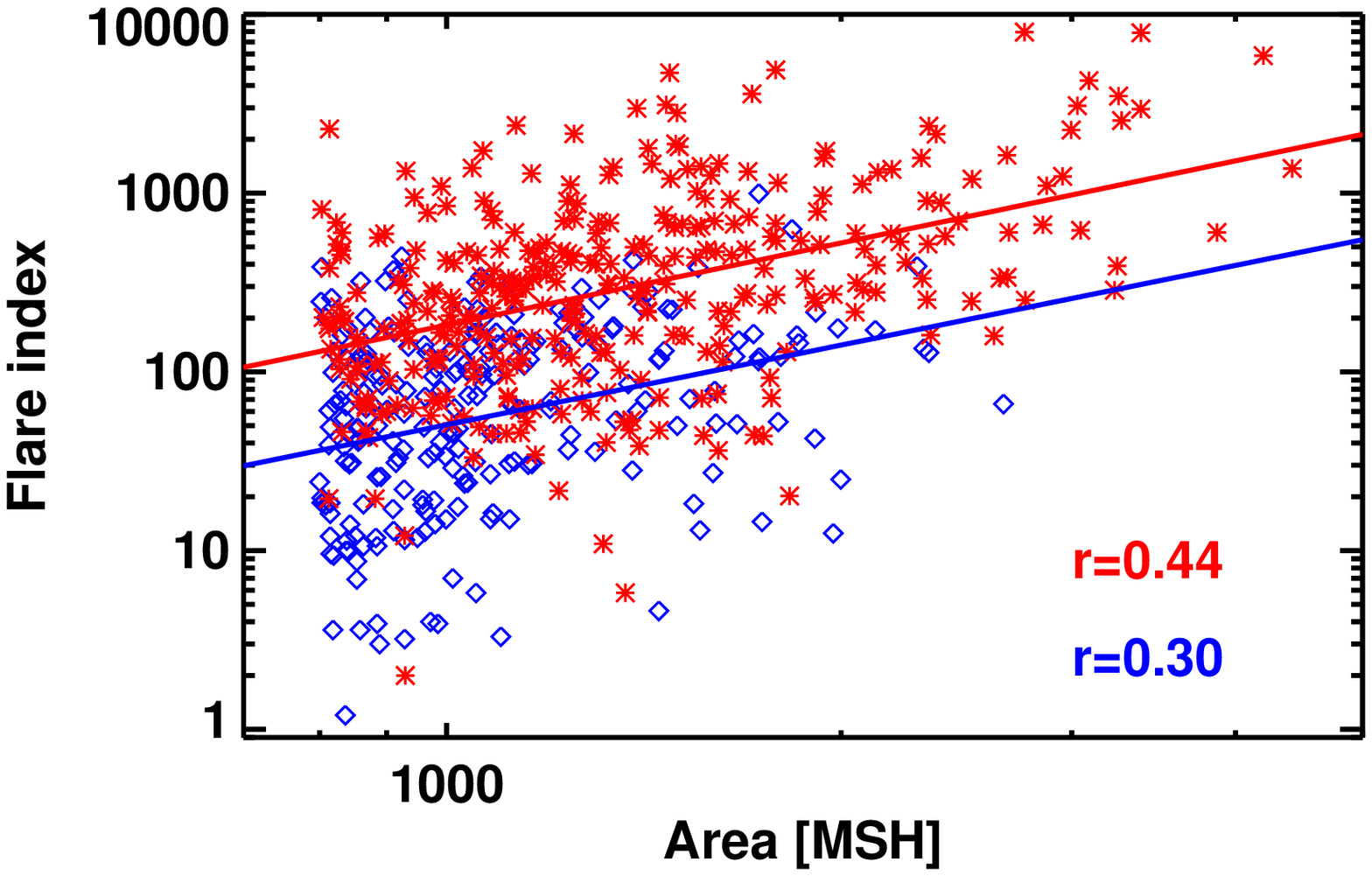}
\includegraphics[scale=0.4]{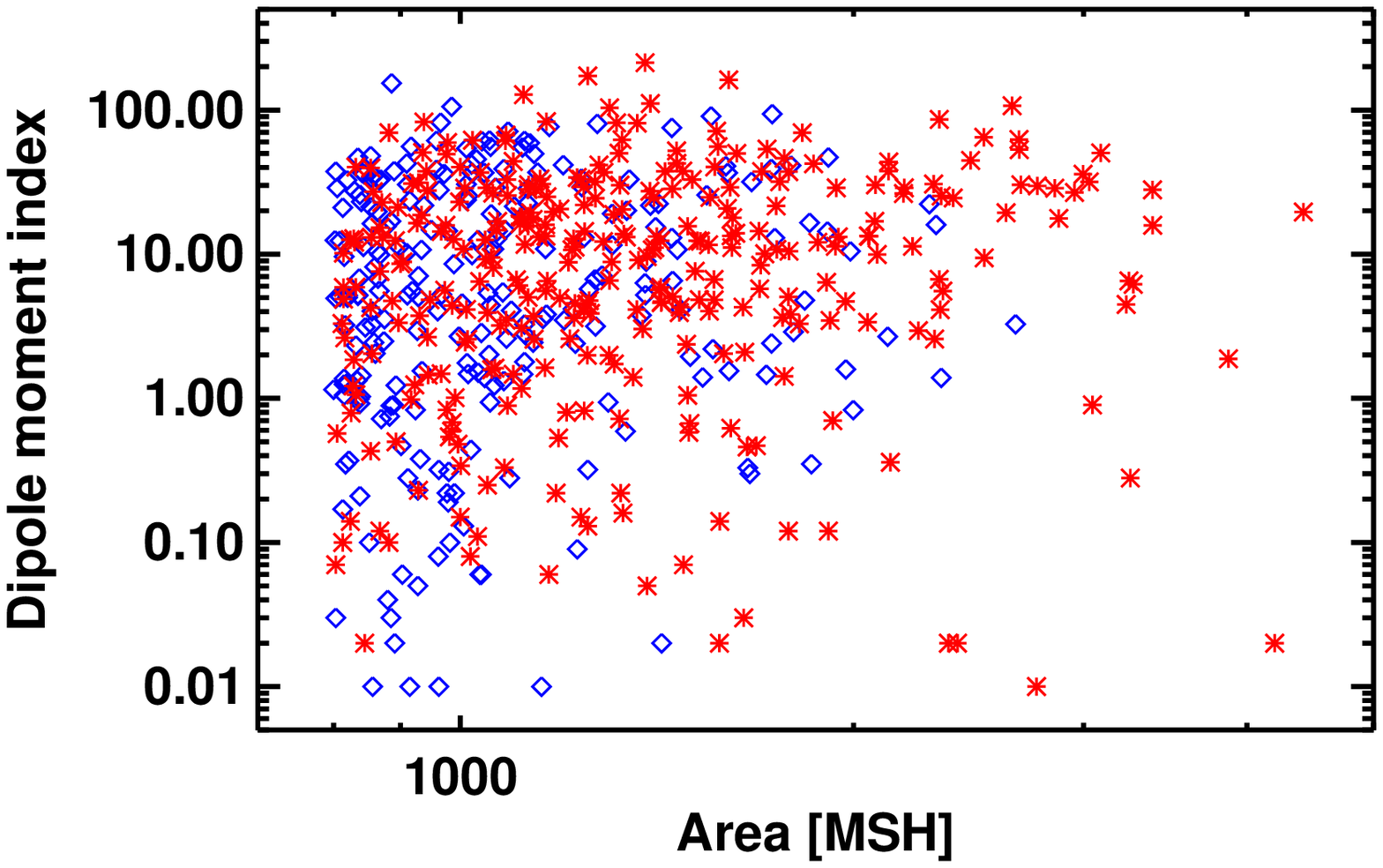}
\caption{Area dependence of the flare index $FI$ (left panel) and axial dipole moment index $DI$ (right panel). Simple ARs are shown by blue diamonds and complex ARs are shown by red asterisks. The red and blue lines in the left panel are the curve fits between the area and the flare index.} \label{fig:areaVSflarDM}
\end{center}
\end{figure}

Figure \ref{fig:areaVSflarDM} is the area dependence of the flare index $FI$ and the dipole moment index $DI$ for both simple and complex ARs. Complex ARs show a stronger correlation ($r=0.44$) between the size of the ARs and the $FI$ values than simple ARs ($r=0.30$). Larger complex ARs tend to have larger $FI$, which is consistent with \cite{Sammis2000}. The curve fittings for the correlation between the size of ARs and the $FI$ are $FI=10^{-2.31}A^{1.53}$ (complex ARs) and $FI=10^{-2.74}A^{1.48}$ (simple ARs). The power-law indices are close to the results of \cite{Takizawa2015}, who analyzed ARs in cycle 23. By contrast, there are no correlations between the size of ARs and the $DI$ for both the simple and complex ARs although the $DI$ value is proportional to the area based on Eq.(\ref{eq:di}). This results from the small area range since only ARs larger than 800$\mu$Hem are selected. The $DI$ values are dominated by latitudes and tilt angles, causing a large range of $DI$ values for a given AR area.

\section{Summary and discussion}
\label{sec:summary}
This paper aims to suggest that different exceptional ARs have different contributions to space weather and space climate. We note that simple ARs, which are usually flare-poor, have the same possibility as complex ARs, which are usually flare-rich, to significantly contribute to the long-term variation of the axial dipole moment and hence to space climate. We take ARs 12673 and 12674 as examples to demonstrate this idea. A statistical analysis of 567 ARs larger than 800$\mu$Hem from 1976 to 2017 is also presented. We suggest two proxies, i.e., the flare index $FI$ and the axial dipole moment index $DI$, to quantify the impacts of ARs on space weather and space climate, respectively. The strong space weather effects imposed by ARs, e.g., the $\delta$-type AR 12673, mainly depend on the magnetic complexity of ARs. The strong space climate effects imposed by ARs mainly depend on the latitudinal location and the latitudinal separation of the positive and negative fluxes. ARs with different degrees of complexity have the same possibility to affect space climate. Furthermore, the SFT simulation assimilating the isolated AR 12673 indicates that it is an AR with abnormal polarity, which weakens the axial dipole moment of cycle 24.

The study of different exceptional ARs affecting short-term and long-term solar variability sheds new light on the studies of stellar magnetic activities and stellar magnetic cycles, especially from the perspective of superflares and magnetic field topology. As a byproduct of our study, we have verified the results of \cite{Sammis2000} based on our 41 yr samples, which include a much longer dataset than the 8 yr data set of \cite{Sammis2000}. Indeed, magnetic complexity plays a more important role in producing severe flares than magnetic flux (area). However, there are arguments based on both observations \citep{Notsu2013,Candelaresi2014} and the modelling \citep{Shibata2013,Kitchatinov2016} that the super-flare energy depends on the total magnetic energy or the dynamo action. Magnetic complexity, which is relevant to the flux emergence process \citep{Fang2015, Chatterjee2016} and the magnetic field topology of the stars \citep{Morin2008, Petit2008,See2016}, is usually ignored. Several surveys have uncovered the magnetic field topologies of stars with different ages and spectral types. A nice summary of large-scale magnetic topologies of cool stars can be found in \cite{Lehmann2018}. If the BL mechanism still works for cool stars, the major idea presented in this paper for the Sun will also be applicable for short-term and long-term stellar activity variations. Big starspot groups in simple bipolar configurations have a weak possibility of generating superflares, but they could contribute significantly to the stellar poloidal, and hence the toroidal, field amplitude provided they emerge around low latitudes with big tilts. This is likely the case for fast rotators with a strong Coriolis force. SFT simulations have shown that starspots that emerge at the low latitudes can be transported to the poles to form the stellar polar spots implied by observations \citep{Schrijver2001,Mackay2004,Isik2011}.

\begin{acknowledgements}
The authors thank the anonymous referee for useful comments and suggestions, which improved the paper. T\"{u}nde Baranyi, the head of the Debrecen solar physics group of Konkoly Observatory, passed away on 24 July 2018. The sunspot databases she led are essential resources for the understanding of the solar cycle. We extend our appreciation to Gopal Hazra and Manfred Sch\"{u}ssler for the helpful comments on this paper. The SDO/HMI data are courtesy of NASA and the SDO/HMI team. We acknowledge the support by the National Science Foundation of China (grant Nos. 11522325, 11873023, 41404136, and 11573038) and by the Fundamental Research Funds for the Central Universities of China.
\end{acknowledgements}


\begin{thebibliography}{}
\bibitem[{{Abramenko}(2005)}]{Abramenko2005}
{Abramenko}, V.~I. 2005, \apj, 629, 1141

\bibitem[{{Antalova}(1996)}]{Antalova1996}
{Antalova}, A. 1996, Contributions of the Astronomical Observatory Skalnate
  Pleso, 26, 98

\bibitem[{{Babcock}(1961)}]{Babcock1961}
{Babcock}, H.~W. 1961, \apj, 133, 572

\bibitem[{{Baranyi}(2015)}]{Baranyi2015}
{Baranyi}, T. 2015, \mnras, 447, 1857

\bibitem[{{Baranyi} {et~al.}(2016){Baranyi}, {Gy{\H o}ri}, \&
  {Ludm{\'a}ny}}]{Baranyi2016}
{Baranyi}, T., {Gy{\H o}ri}, L., \& {Ludm{\'a}ny}, A. 2016, \solphys, 291, 3081

\bibitem[{{Baranyi} {et~al.}(2001){Baranyi}, {Gyori}, {Ludm{\'a}ny}, \&
  {Coffey}}]{Baranyi2001}
{Baranyi}, T., {Gyori}, L., {Ludm{\'a}ny}, A., \& {Coffey}, H.~E. 2001, \mnras,
  323, 223

\bibitem[{{Baranyi} {et~al.}(2013){Baranyi}, {Kir{\'a}ly}, \&
  {Coffey}}]{Baranyi2013}
{Baranyi}, T., {Kir{\'a}ly}, S., \& {Coffey}, H.~E. 2013, \mnras, 434, 1713

\bibitem[{{Baumann} {et~al.}(2004){Baumann}, {Schmitt}, {Sch{\"u}ssler}, \&
  {Solanki}}]{Baumann2004}
{Baumann}, I., {Schmitt}, D., {Sch{\"u}ssler}, M., \& {Solanki}, S.~K. 2004,
  \aap, 426, 1075

\bibitem[{{Cameron} {et~al.}(2016){Cameron}, {Jiang}, \&
  {Sch{\"u}ssler}}]{Cameron2016}
{Cameron}, R.~H., {Jiang}, J., \& {Sch{\"u}ssler}, M. 2016, \apjl, 823, L22

\bibitem[{{Cameron} \& {Sch{\"u}ssler}(2015)}]{Cameron2015}
{Cameron}, R.~H., \& {Sch{\"u}ssler}, M. 2015, Science, 347, 1333

\bibitem[{{Candelaresi} {et~al.}(2014){Candelaresi}, {Hillier}, {Maehara},
  {Brandenburg}, \& {Shibata}}]{Candelaresi2014}
{Candelaresi}, S., {Hillier}, A., {Maehara}, H., {Brandenburg}, A., \&
  {Shibata}, K. 2014, \apj, 792, 67

\bibitem[{{Chatterjee} {et~al.}(2016){Chatterjee}, {Hansteen}, \&
  {Carlsson}}]{Chatterjee2016}
{Chatterjee}, P., {Hansteen}, V., \& {Carlsson}, M. 2016, Physical Review
  Letters, 116, 101101

\bibitem[{{Chen} \& {Wang}(2012)}]{Chen2012}
{Chen}, A.~Q., \& {Wang}, J.~X. 2012, \aap, 543, A49

\bibitem[{{Chen} {et~al.}(2011){Chen}, {Wang}, {Li}, {Feynman}, \&
  {Zhang}}]{Chen2011}
{Chen}, A.~Q., {Wang}, J.~X., {Li}, J.~W., {Feynman}, J., \& {Zhang}, J. 2011,
  \aap, 534, A47

\bibitem[{{Fang} \& {Fan}(2015)}]{Fang2015}
{Fang}, F., \& {Fan}, Y. 2015, \apj, 806, 79

\bibitem[{{Gopalswamy}(2006)}]{Gopalswamy2006}
{Gopalswamy}, N. 2006, Journal of Astrophysics and Astronomy, 27, 243

\bibitem[{{Gopalswamy} {et~al.}(2004){Gopalswamy}, {Nunes}, {Yashiro}, \&
  {Howard}}]{Gopalswamy2004}
{Gopalswamy}, N., {Nunes}, S., {Yashiro}, S., \& {Howard}, R.~A. 2004, Advances
  in Space Research, 34, 391

\bibitem[{{Guo} {et~al.}(2014){Guo}, {Lin}, \& {Deng}}]{Guo2014}
{Guo}, J., {Lin}, J., \& {Deng}, Y. 2014, \mnras, 441, 2208

\bibitem[{{Gy{\H o}ri} {et~al.}(2017){Gy{\H o}ri}, {Ludm{\'a}ny}, \&
  {Baranyi}}]{Gyori2017}
{Gy{\H o}ri}, L., {Ludm{\'a}ny}, A., \& {Baranyi}, T. 2017, \mnras, 465, 1259

\bibitem[{{I{\c s}{\i}k} {et~al.}(2011){I{\c s}{\i}k}, {Schmitt}, \&
  {Sch{\"u}ssler}}]{Isik2011}
{I{\c s}{\i}k}, E., {Schmitt}, D., \& {Sch{\"u}ssler}, M. 2011, \aap, 528, A135

\bibitem[{{Jaeggli} \& {Norton}(2016)}]{Jaeggli2016}
{Jaeggli}, S.~A., \& {Norton}, A.~A. 2016, \apjl, 820, L11

\bibitem[{{Jiang} {et~al.}(2013){Jiang}, {Cameron}, {Schmitt}, \& {I{\c
  s}{\i}k}}]{Jiang2013}
{Jiang}, J., {Cameron}, R.~H., {Schmitt}, D., \& {I{\c s}{\i}k}, E. 2013, \aap,
  553, A128

\bibitem[{{Jiang} {et~al.}(2014{\natexlab{a}}){Jiang}, {Cameron}, \&
  {Sch{\"u}ssler}}]{Jiang2014b}
{Jiang}, J., {Cameron}, R.~H., \& {Sch{\"u}ssler}, M. 2014{\natexlab{a}}, \apj,
  791, 5

\bibitem[{{Jiang} {et~al.}(2015){Jiang}, {Cameron}, \&
  {Sch{\"u}ssler}}]{Jiang2015}
---. 2015, \apjl, 808, L28

\bibitem[{{Jiang} \& {Cao}(2018)}]{Jiang2018a}
{Jiang}, J., \& {Cao}, J. 2018, Journal of Atmospheric and Solar-Terrestrial
  Physics, 176, 34

\bibitem[{{Jiang} {et~al.}(2014{\natexlab{b}}){Jiang}, {Hathaway}, {Cameron},
  {Solanki}, {Gizon}, \& {Upton}}]{Jiang2014a}
{Jiang}, J., {Hathaway}, D.~H., {Cameron}, R.~H., {et~al.} 2014{\natexlab{b}},
  \ssr, 186, 491

\bibitem[{{Jiang} {et~al.}(2010){Jiang}, {I{\c s}ik}, {Cameron}, {Schmitt}, \&
  {Sch{\"u}ssler}}]{Jiang2010}
{Jiang}, J., {I{\c s}ik}, E., {Cameron}, R.~H., {Schmitt}, D., \&
  {Sch{\"u}ssler}, M. 2010, \apj, 717, 597

\bibitem[{{Jiang} {et~al.}(2018){Jiang}, {Wang}, {Jiao}, \& {Cao}}]{Jiang2018b}
{Jiang}, J., {Wang}, J.-X., {Jiao}, Q.-R., \& {Cao}, J.-B. 2018, \apj, 863, 159

\bibitem[{{Jing} {et~al.}(2006){Jing}, {Song}, {Abramenko}, {Tan}, \&
  {Wang}}]{Jing2006}
{Jing}, J., {Song}, H., {Abramenko}, V., {Tan}, C., \& {Wang}, H. 2006, \apj,
  644, 1273

\bibitem[{{Kilpua} {et~al.}(2015){Kilpua}, {Lumme}, {Andreeova}, {Isavnin}, \&
  {Koskinen}}]{Kilpua2015}
{Kilpua}, E.~K.~J., {Lumme}, E., {Andreeova}, K., {Isavnin}, A., \& {Koskinen},
  H.~E.~J. 2015, Journal of Geophysical Research (Space Physics), 120, 4112

\bibitem[{{Kitchatinov} \& {Olemskoy}(2016)}]{Kitchatinov2016}
{Kitchatinov}, L.~L., \& {Olemskoy}, S.~V. 2016, \mnras, 459, 4353

\bibitem[{{K{\"u}nzel}(1960)}]{Kunzel1960}
{K{\"u}nzel}, H. 1960, Astronomische Nachrichten, 285, 271

\bibitem[{{Lehmann} {et~al.}(2018){Lehmann}, {Jardine}, {Mackay}, \&
  {Vidotto}}]{Lehmann2018}
{Lehmann}, L.~T., {Jardine}, M.~M., {Mackay}, D.~H., \& {Vidotto}, A.~A. 2018,
  \mnras, 478, 4390

\bibitem[{{Leighton}(1969)}]{Leighton1969}
{Leighton}, R.~B. 1969, \apj, 156, 1

\bibitem[{{Li} \& {Ulrich}(2012)}]{Li2012}
{Li}, J., \& {Ulrich}, R.~K. 2012, \apj, 758, 115

\bibitem[{{Liu} \& {Zhang}(2002)}]{Liu2002}
{Liu}, Y., \& {Zhang}, H.~Q. 2002, \aap, 386, 646

\bibitem[{{Liu} {et~al.}(2012){Liu}, {Hoeksema}, {Scherrer}, {Schou},
  {Couvidat}, {Bush}, {Duvall}, {Hayashi}, {Sun}, \& {Zhao}}]{Liu2012}
{Liu}, Y., {Hoeksema}, J.~T., {Scherrer}, P.~H., {et~al.} 2012, \solphys, 279,
  295

\bibitem[{{Mackay} {et~al.}(2004){Mackay}, {Jardine}, {Collier Cameron},
  {Donati}, \& {Hussain}}]{Mackay2004}
{Mackay}, D.~H., {Jardine}, M., {Collier Cameron}, A., {Donati}, J.-F., \&
  {Hussain}, G.~A.~J. 2004, \mnras, 354, 737

\bibitem[{{Mackay} {et~al.}(2002){Mackay}, {Priest}, \&
  {Lockwood}}]{Mackay2002}
{Mackay}, D.~H., {Priest}, E.~R., \& {Lockwood}, M. 2002, \solphys, 207, 291

\bibitem[{{Mackay} \& {Yeates}(2012)}]{Mackay2012}
{Mackay}, D.~H., \& {Yeates}, A.~R. 2012, Living Reviews in Solar Physics, 9, 6

\bibitem[{{Morin} {et~al.}(2008){Morin}, {Donati}, {Petit}, {Delfosse},
  {Forveille}, {Albert}, {Auri{\`e}re}, {Cabanac}, {Dintrans}, {Fares},
  {Gastine}, {Jardine}, {Ligni{\`e}res}, {Paletou}, {Ramirez Velez}, \&
  {Th{\'e}ado}}]{Morin2008}
{Morin}, J., {Donati}, J.-F., {Petit}, P., {et~al.} 2008, \mnras, 390, 567

\bibitem[{{Mursula} {et~al.}(2007){Mursula}, {Usoskin}, \&
  {Maris}}]{Mursula2007}
{Mursula}, K., {Usoskin}, I.~G., \& {Maris}, G. 2007, Advances in Space
  Research, 40, 885

\bibitem[{{Nagy} {et~al.}(2017){Nagy}, {Lemerle}, {Labonville}, {Petrovay}, \&
  {Charbonneau}}]{Nagy2017}
{Nagy}, M., {Lemerle}, A., {Labonville}, F., {Petrovay}, K., \& {Charbonneau},
  P. 2017, \solphys, 292, 167

\bibitem[{{Nandy} \& {Martens}(2007)}]{Nandy2007}
{Nandy}, D., \& {Martens}, P.~C.~H. 2007, Advances in Space Research, 40, 891

\bibitem[{{Notsu} {et~al.}(2013){Notsu}, {Shibayama}, {Maehara}, {Notsu},
  {Nagao}, {Honda}, {Ishii}, {Nogami}, \& {Shibata}}]{Notsu2013}
{Notsu}, Y., {Shibayama}, T., {Maehara}, H., {et~al.} 2013, \apj, 771, 127

\bibitem[{{Petit} {et~al.}(2008){Petit}, {Dintrans}, {Solanki}, {Donati},
  {Auri{\`e}re}, {Ligni{\`e}res}, {Morin}, {Paletou}, {Ramirez Velez},
  {Catala}, \& {Fares}}]{Petit2008}
{Petit}, P., {Dintrans}, B., {Solanki}, S.~K., {et~al.} 2008, \mnras, 388, 80

\bibitem[{{Romano} \& {Zuccarello}(2007)}]{Romano2007}
{Romano}, P., \& {Zuccarello}, F. 2007, \aap, 474, 633

\bibitem[{{Sammis} {et~al.}(2000){Sammis}, {Tang}, \& {Zirin}}]{Sammis2000}
{Sammis}, I., {Tang}, F., \& {Zirin}, H. 2000, \apj, 540, 583

\bibitem[{{Schrijver} \& {Title}(2001)}]{Schrijver2001}
{Schrijver}, C.~J., \& {Title}, A.~M. 2001, \apj, 551, 1099

\bibitem[{{Seaton} \& {Darnel}(2018)}]{Seaton2018}
{Seaton}, D.~B., \& {Darnel}, J.~M. 2018, \apjl, 852, L9

\bibitem[{{See} {et~al.}(2016){See}, {Jardine}, {Vidotto}, {Donati}, {Boro
  Saikia}, {Bouvier}, {Fares}, {Folsom}, {Gregory}, {Hussain}, {Jeffers},
  {Marsden}, {Morin}, {Moutou}, {do Nascimento}, {Petit}, \& {Waite}}]{See2016}
{See}, V., {Jardine}, M., {Vidotto}, A.~A., {et~al.} 2016, \mnras, 462, 4442

\bibitem[{{Shen} {et~al.}(2018){Shen}, {Xu}, {Wang}, {Chi}, \&
  {Luo}}]{Shen2018}
{Shen}, C., {Xu}, M., {Wang}, Y., {Chi}, Y., \& {Luo}, B. 2018, \apj, 861, 28

\bibitem[{{Shi} \& {Wang}(1994)}]{Shi1994}
{Shi}, Z., \& {Wang}, J. 1994, \solphys, 149, 105

\bibitem[{{Shibata} \& {Magara}(2011)}]{Shibata2011}
{Shibata}, K., \& {Magara}, T. 2011, Living Reviews in Solar Physics, 8, 6

\bibitem[{{Shibata} {et~al.}(2013){Shibata}, {Isobe}, {Hillier}, {Choudhuri},
  {Maehara}, {Ishii}, {Shibayama}, {Notsu}, {Notsu}, {Nagao}, {Honda}, \&
  {Nogami}}]{Shibata2013}
{Shibata}, K., {Isobe}, H., {Hillier}, A., {et~al.} 2013, \pasj, 65, 49

\bibitem[{{Stenflo} \& {Kosovichev}(2012)}]{Stenflo2012}
{Stenflo}, J.~O., \& {Kosovichev}, A.~G. 2012, \apj, 745, 129

\bibitem[{{Takizawa} \& {Kitai}(2015)}]{Takizawa2015}
{Takizawa}, K., \& {Kitai}, R. 2015, \solphys, 290, 2093

\bibitem[{{Toriumi} {et~al.}(2017){Toriumi}, {Schrijver}, {Harra}, {Hudson}, \&
  {Nagashima}}]{Toriumi2017}
{Toriumi}, S., {Schrijver}, C.~J., {Harra}, L.~K., {Hudson}, H., \&
  {Nagashima}, K. 2017, \apj, 834, 56

\bibitem[{{Upton} \& {Hathaway}(2014)}]{Upton2014}
{Upton}, L., \& {Hathaway}, D.~H. 2014, \apj, 780, 5

\bibitem[{{van Ballegooijen} {et~al.}(1998){van Ballegooijen}, {Cartledge}, \&
  {Priest}}]{Ballegooijen1998}
{van Ballegooijen}, A.~A., {Cartledge}, N.~P., \& {Priest}, E.~R. 1998, \apj,
  501, 866

\bibitem[{{Wang} {et~al.}(1989){Wang}, {Nash}, \& {Sheeley}}]{Wang1989b}
{Wang}, Y.-M., {Nash}, A.~G., \& {Sheeley}, Jr., N.~R. 1989, Science, 245, 712

\bibitem[{{Wang} \& {Sheeley}(1991)}]{Wang1991}
{Wang}, Y.-M., \& {Sheeley}, Jr., N.~R. 1991, \apj, 375, 761

\bibitem[{{Wang} {et~al.}(2000){Wang}, {Sheeley}, \& {Lean}}]{Wang2000}
{Wang}, Y.-M., {Sheeley}, Jr., N.~R., \& {Lean}, J. 2000, \grl, 27, 621

\bibitem[{{Whitbread} {et~al.}(2018){Whitbread}, {Yeates}, \&
  {Mu{\~n}oz-Jaramillo}}]{Whitbread2018}
{Whitbread}, T., {Yeates}, A.~R., \& {Mu{\~n}oz-Jaramillo}, A. 2018, ArXiv
  e-prints, arXiv:1807.01617

\bibitem[{{Yan} {et~al.}(2018){Yan}, {Wang}, {Pan}, {Kong}, {Xue}, {Yang},
  {Li}, \& {Feng}}]{Yan2018}
{Yan}, X.~L., {Wang}, J.~C., {Pan}, G.~M., {et~al.} 2018, \apj, 856, 79

\bibitem[{{Yang} {et~al.}(2017{\natexlab{a}}){Yang}, {Zhang}, {Zhu}, \&
  {Song}}]{YangS2017}
{Yang}, S., {Zhang}, J., {Zhu}, X., \& {Song}, Q. 2017{\natexlab{a}}, \apjl,
  849, L21

\bibitem[{{Yang} {et~al.}(2017{\natexlab{b}}){Yang}, {Hsieh}, {Yu}, \&
  {Chen}}]{YangY2017}
{Yang}, Y.-H., {Hsieh}, M.-S., {Yu}, H.-S., \& {Chen}, P.~F.
  2017{\natexlab{b}}, \apj, 834, 150

\bibitem[{{Yeates} {et~al.}(2015){Yeates}, {Baker}, \& {van
  Driel-Gesztelyi}}]{Yeates2015}
{Yeates}, A.~R., {Baker}, D., \& {van Driel-Gesztelyi}, L. 2015, \solphys, 290,
  3189

\bibitem[{{Zirin} \& {Liggett}(1987)}]{Zirin1987}
{Zirin}, H., \& {Liggett}, M.~A. 1987, \solphys, 113, 267

\end{thebibliography}

\clearpage

\end{document}